# An Automated System for Essay Scoring of Online Exams in Arabic based on Stemming Techniques and Levenshtein Edit Operations

Emad Fawzi Al-Shalabi

**Department of Information Technology, AL-BALQA Applied University**
**Al-Huson University College, Irbid, Al-Huson, 50, Jordan**

**Abstract**

In this article, an automated system is proposed for essay scoring in Arabic language for online exams based on stemming techniques and Levenshtein edit operations. An online exam has been developed on the proposed mechanisms, exploiting the capabilities of light and heavy stemming. The implemented online grading system has shown to be an efficient tool for automated scoring of essay questions.

## 1. Introduction

Recently, automated grading systems have gained increasing attention due to their convenience over traditional grading methods. The rapidly growing reliance on technology in the educational field [1] and the increasing numbers of students; raised the need for an efficient scoring mechanism to fully replace the teacher's role in the scoring process, while saving time and guaranteeing fairness. The process of Automatic Scoring (AS) addresses the evaluation of a student's answer by performing a comparison with a model answer.

Several types of tools and web-services have been developed with the purpose of performing automatic scoring of online exams with little or no user intervention. One of the easiest ways to implement automatic scoring is through the adoption of multiple choice (MC) exams [2], due to their nature and ease in scoring by a computer; MC exams have become widely used in online exams. Despite the presented advantages, MC question format has been criticized of unfairness, because they allow students to pick the correct answer based on chance, rendering it difficult to distinguish between a student who chose the correct answer based on exam preparation and the understanding of the presented problem and another who blindly guessed the answer. The case is also similar for true-false and matching question formats.

On the other hand, essay questions present far superior advantages over the previously mentioned question formats. Essay questions can reflect the depth of a student's knowledge and problem solving skills; they can also provide feedback for the instructor by shedding some light on the student's erroneous conclusions.

Another advantage of implementing AES systems is to remove the subjectivity in traditional scoring methods, where the instructor grades essay questions based on their own interpretation of a given answer. This as a result, ensures that a standardized basis for question scoring is being applied for all students alike. However, the implementation of Automatic Essay Scoring (AES) mechanisms is a rather difficult task in comparison with MC-based AS systems, this is mainly because essay answers a complicated process of text analysis.

Several AES models have been developed since the 1960s, and due to the growing use of technology in the educational system in past decade, AES has become a very important area in the research field. However, the majority of available research is more concerned with automated scoring for English language essay questions. However, there is a lack of research when it comes to AES mechanisms for other languages such as Arabic, despite being a widely used language, which raises the important of investigating new mechanisms of automated scoring for essay question.

In this research, we propose a stemming-based mechanism for automatic essay scoring in Arabic language. This paper is organized as follows:

Section II presents related works on automated essay scoring systems for Arabic, Section III introduces the problem statement explaining the challenges in Arabic language processing, in Section IV an automatic essay scoring mechanism is proposed, Section V presents the experimental work produced based on the proposed mechanism, Section VI shows the conclusions of the research and future work.

## 2. Related Works

One of the most commonly used AES models is presented in [3] other AES models are [2] and [1]

However, it is highly difficult to implement the mentioned mechanisms for Arabic language, due to its complex nature, being highly inflectional and ambiguous in the absence of diacritics. There have been only few attempts in research on this subject, and so far none of them has been able to provide a fully functional auto-grading system.

In [4], the authors proposed an Arabic web-based examination system, where students can login using a username and password to take exams online, exam questions are stored along with correct answers on the server, answers are auto-graded by performing a comparison between the correct answer and the student's answer. However, the proposed system did not provide auto-grading for essay questions, answers to such questions are sent to the instructor for manual grading then passed back to the system, this long and complex process renders the system rather impractical, prone to error and misuse.

Text similarity techniques were used in [3] for the purpose of short answer auto-grading in Arabic language. The article presents an evaluation of the effect of combining corpus-based and string-based similarity measures. Feedback is also provided to students during the exam through comments that describe the answer's level of correctness. However, the system also requires human intervention and is not fully automated.

In [Khalid], automatic grading for online exams is proposed using statistical and computational linguistic techniques, where a variety of statistical distributions are employed to give weights to the words of the instructor's answer, utilizing human-computer interaction to benefit the grading system.

## 3. Problem Statement

It is clear from the previous section, that there is a lack in the number of research concerned with automated grading of essay questions in Arabic language, the majority of available research available does not provide full implementation of the proposed techniques and often requires manual grading by an instructor at some point

An efficient automated grading system for essay questions in Arabic language should prove the possess the following features:

- Fully automated grading capabilities without the requirement of human intervention
- Low in computational complexity to allow for fast grading and lightweight implementation for web-services complying the nature of online exams.
- Efficient handling of the various complex aspects in Arabic language.

This article aims to solve the problem in hand by developing an automated essay scoring mechanism that is both efficient and low in complexity for use in online web-based exams, without the requirement of manual grading.

## 4. Automatic Essay Scoring Mechanism

In this article, stemming techniques were exploited for the purpose of auto-grading essay questions in online exams. A stemming algorithm may be defined as the procedure of reducing all words that share the same stem to a common form [5].

The proposed scoring system is divided into two algorithms; heavy stemming and light stemming, the general structure of the scoring system is show in figure (1) which explains the general mechanism of the system.

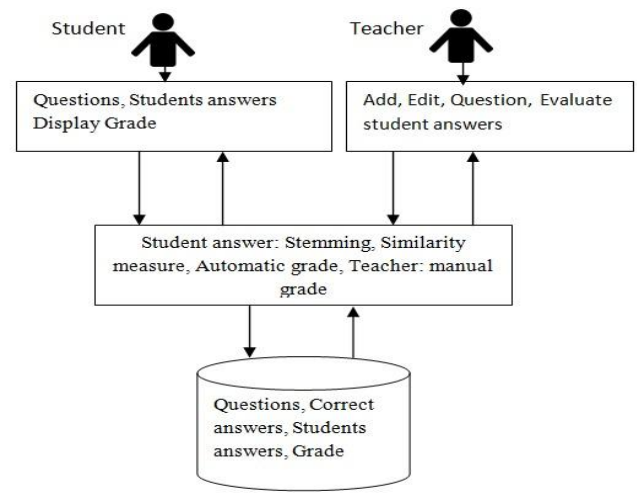

Fig. 1 Automatic essay scoring architecture

Each question is loaded from the database and displayed along with a form for the student to fill in the answer, the student's answer is obtained from the form while the correct answer is retrieved from the database for comparison.

4.1 Heavy Stemming Approach

Heavy stemming, also referred to as root-based stemming begins with removing well-known prefixes and suffixes to extract the actual root of a word, the identifies the pattern in correspondence with the remaining word.

The auto-grading process is carried as follows:

Step 1: Get both question and correct answer from the database.
Step 2: Get the student's answer from the form.

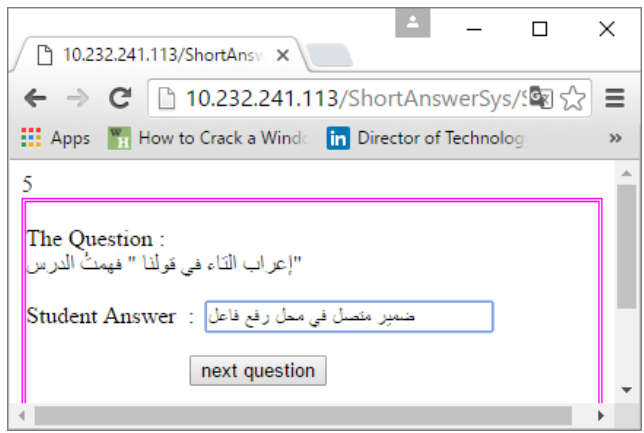

Fig. 2 Answer form

Step 3: Begin heavy stemming on both the student's answer and the correct answer using, this procedure involves three steps:

1. Removal of numbers from both answers.
2. Removal of diacritics from both answers.
3. Removal of any letters from other languages.

Step 4: Split each one of the two answers into an array of words, processing one word at a time as follows:

1- loop through words of each answer and remove stop words , a list of stop words is available in database ( في , و , ان , اذا , هو , هي , هما )
2- remove the (AL) , and its Derivatives , also available in database ( بال , لل , ال , فال , لبال , وبال , فبال , تال , وال , كال ,)
3- Normalize words by replacing similar letters ( إ,آ,أ with ا, ة with ه )
4- Remove prefix if word length is greater than 3 , else skip this step
5- Remove suffix , if word length is greater than 3 , else skip this step , note that on the case of heavy stemming a different list of suffixes is provided

Step 5: Find the similarities by giving a weight to each word in both answers.

This step requires finding the edit distance, which can be obtained following the two Eq.(1,2):
1- The edit distance, which is the minimum number of operations required to transform one string into another.
$$D(s_1, s_2) \qquad (1)$$

2- The similarity equation:
$$S(s_1, s_2) = \frac{1 - D(s_1, s_2)}{Max(L(s_1), L(s_2))} \qquad (2)$$

Where L is the length of a given string.

Step 6: Set weight for each word by:
$$Word_i\ Weight = \frac{1}{Total\ words\ in\ Correct\ answer} \qquad (3)$$

Step 7: For each word in student answer calculate the similarity with words in correct answer:

A. If similarity between $StudentWord_i$ and $CorrectWord_i = 1$ then add weight to the final mark.
$$FinalMark\mathrel{+}= Word_i\ Weight \qquad (4)$$
B. Else, if the similarity between $StudentWord_i$ and $CorrectWord_i < 1$ and $>= 0.96$, add weight to the final mark using Eq. (4).
Note that if the similarity is greater than or equal to 0.96, then it is considered a correct word, this percentage can be changed by the instructor.
C. Else, if the similarity between $StudentWord_i$ and $CorrectWord_i$ is $< 0.96$ and $>= 0.80$ then add half the weight to the final mark.
Note that if the similarity is less than 0.96 then it is considered an incomplete answer/word, this percentage can be decided by the instructor.
$$FinalMark\mathrel{+}= Word_i Weight * 0.5$$
D. Else, if the similarity between $StudentWord_i$ and $CorrectWord_i$ is $< 0.80$ then no weight is added to the final mark.
Note that if the similarity is less than 0.80 then it is considered a wrong answer, this percentage can also be changed by the instructor.
$$FinalMark\mathrel{+}= 0$$
E. Display the final mark, that is, the sum of weights for each word in the student's answer.
F. Move to next question, if there's one.

4.2 Light Stemming Approach

Light stemming is rather a less complex process, where the stemming is stopped upon the removal of prefixes and suffixes, without attempting to identify the actual root of the word.

The auto-grading process is carried as follows:

Step 1: Get both question and correct answer from the database.

Step 2: Get the student's answer from the form.

Step 3: Begin the stemming process on both the correct answer and the student's answer as follows:

1- Remove numbers from both answers.
2- Remove letters from other languages (i.e. English).

Step 4: Split each of the two answers into an array, processed one word at a time as follows:

1- loop through words of each answer and remove stop words. A list of stop words is available in the database (في , و , ان , اذا , هو , هي , هما ,)
2- Remove the (AL), and its Derivatives available in the databse: ( , بال , لل , ال , فال , لبال , وبال , فبال , تال , وال , كال)
3- Normalize words by replacing similar letters.( أ,آ,إwith  ا), (ة with ه).
4- Remove suffixes if word length is greater than 3. Else, skip this step.
   Note that in the case of light stemming, a different list of suffixes is provided, including 10 suffixes.

Step 5: Finding similarity, this is done by giving each word in both answers a weight, which requires finding the edit distance between the two words using Eq. (1) and Eq. (2) respectively.

A. Set the weight of each word using Eq. (3)
B. For each word in student answer calculate the similarity with words in correct answer:

C. If similarity between $StudentWord_i$ and $CorrectWord_i = 1$ then add weight to the final mark using Eq. (4).

D. Else, if the similarity between $StudentWord_i$ and $CorrectWord_i < 1$ and $>= 0.96$, add weight to the final mark using Eq. (4).

Note that if the similarity is greater than or equal to 0.96, then it is considered a correct word, this percentage can be changed by the instructor.

E. Else, if the similarity between $StudentWord_i$ and $CorrectWord_i$ is $< 0.96$ and $>= 0.8$ then add half the weight to the final mark.
Note that if the similarity is less than 0.96 then it is considered an incomplete answer/word, this percentage can be decided by the instructor.
$$FinalMark += Word_i Weight * 0.5$$

F. Else, if the similarity between $StudentWord_i$ and $CorrectWord_i$ is less than 0.80 then no weight is added to the final mark.
Note that if the similarity is less than 0.80 then it is considered a wrong answer, this percentage can also be changed by the instructor.
$$FinalMark += 0$$

G. Display the final mark, that is, the sum of weights for each word in the student's answer.
H. Move to next question, if there's one.

## 5. Experimental Work

A web-service has been developed based on the proposed scoring mechanisms, an online exam has been conducted to check the efficiency of both mechanisms, and following are two examples demonstrating the automated scoring process.

Example 1:

Question: اكمل البيت التالي ... أقول لأصحابي ارفعوني فإنني

Correct Answer: يقر بعيني أن سهيل بدا ليا

Student Answer: يقر بعيني ان سهيل بدا

Solution:

1. Split each answer into words.
2. Apply normalization, remove stop words, prefixes and suffixes from both answers as long as $word_i$ length $> 3$
3. Find similarity for each word in student answer, demonstrated word by word as follows:
4. Find weight per word = 1/number of words in correct answer = 1/5 = 0.2, this will be the weight for each word.
5. The word يقر from student answer,

For each word in correct answer calculate the similarity, between يقر from student answer and the one in correct answer is 1.

D (يقر, يقر) = 0

S (يقر, يقر) = 1 – 0 / Max (Length (يقر) , Length (يقر))

S (يقر, يقر) = 1 – 0 / 3

S (يقر, يقر) = 1

So MarkSum += weightof word

MarkSum +=0.2

6. The word عين also have a similarity = 1 with عين in correct answer.
   So MarkSum +=0.2, which means MarkSum now = 0.4
7. The word هيل also have a similar word which means MarkSum +=0.2, which means MarkSum now = 0.6
8. The word بدا also have a similar word which means MarkSum +=0.2, which means MarkSum now = 0.8
9. The word ليا in the correct answer have a weight a 0.2 but does not exist in student answer so MarkSum +=0, which means MarkSum still as 0.8
10. Give the result depending on the mark sum that is the sum of all weights MarkSum = 0.8

NOTE the conditions can be changed: if MarkSum = 1 then it's a full mark

Else if MarkSum < 1 & >= 0.96 then it's also considered a correct answer.

Else if MarkSum >= 0.75 & < 0.96 then the student answer might be correct must be checked

Else if MarkSum is less than 0.75 then it's a wrong answer

Here Since MarkSum = 0.8 then the student answer might be correct must be checked

Note that here also both light and heavy scoring systems gave the same answer.

Example 2 :

Question: أحد الآتية غير صحيح فيما يتعلق بالنظرة الإسلامية للقدر

Correct answer: الإيمان لا يوجب العمل

Student Answer: دايما الايمان يوجب العمل

Solution:

1. Split each answer into words
2. Stop words removed, normalized, prefix and suffix removed from both answers as long as word. length > 3
   Correct answer: ايم لا يوجب عمل
   Student answer: دايما ايم يوجب عمل
   NOTE that any additional words in student answer is dropped, in this example the word دايما has no effect
3. Find similarity for each word in student answer, the process is demonstrated word by word, as follows:
4. find weight per word a 1/number of words in correct answer = 1/4 = 0.25, this will be the weight of each word
5. The word ايم from student answer , that is الإيمان after removing ال and suffix ان
6. For each word in correct answer calculate the similarity, the similarity between ايم from student answer and the one in correct answer is 1.

   D ( ايم , ايم ) = 0
   S ( ايم , ايم ) = 1 – 0 / Max (Length ( ايم ) , Length ( ايم ))
   S ( ايم , ايم ) = 1 – 0 / 3
   S ( ايم , ايم ) = 1

   So MarkSum += weightof word
   MarkSum +=0.25
7. the word يوجب also have a similarity = 1 with يوجب in correct answer So MarkSum +=0.25 , which means MarkSum now = 0.50, the word عمل also have a similar word which means MarkSum +=0.25 , which means MarkSum now = 0.75.
8. The word لا in the correct answer have a weight = 0.25 but does not exist in student answer so MarkSum +=0 , which means MarkSum still = 0.75
9. give the result depending on the mark sum that is the sum of all weights

MarkSum = 0.75

NOTE the conditions can be changed :

If MarkSum = 1 then it's a full mark

Else if MarkSum < 1 & >= .96 then it's also considered a correct answer

Else if MarkSum >=.75 & < 0.96 then the student answer might be correct must be checked

Else if Else if MarkSum is less than 0.75 then it's a wrong answer.

Here Since MarkSum = 0.75 then the student answer might be correct must be checked

## 6. Conclusion

In this article, an automated system for essay scoring of Arabic language was proposed. An online examination web-service has been implemented based of the proposed mechanism. Real-life tests of the implemented system have been conducted, and the proposed mechanisms have shown to be an efficient grading tool for essay questions in Arabic language.

In future, the proposed scoring system may be further developed to serve online mathematical exams, by extending the available mechanisms to include numbers and Latin symbols in the scoring process.

**Author** Emad Fawzi Al-shalabi received the B.S. degree in Information Technology, Management Information Systems from Philadelphia University , Jordan in 2004, and the MSc in Computer Information Systems from Yarmouk University (YU), Jordan in 2009.